**Significant and extremely fast sodium storage in graphite – β''- alumina composites via job-sharing**

Chuanhai Gan [1 #], Chuanlian Xiao [1 #], Yue Zhu [1, 2 *], Sebastian Bette [1],
Bettina V. Lotsch [1], Joachim Maier [1 *]

1. Max Planck Institute for Solid State Research, Stuttgart, Germany
2. School of Materials Science and Engineering, Ocean University of China, Qingdao 266404, China

# These authors contributed equally.

* Corresponding authors: zhuyue@ouc.edu.cn, office-maier@fkf.mpg.de

**Abstract**

One weak link in the applicability of sodium-based batteries is the choice of an appropriate negative electrode material. Here we show that Na storage in composites of the electronic conductor graphite and the fast ion conductor Na-β''-$Al_2O_3$ allows for significant storage even though none of the constituent phases does this to a noteworthy degree. The achieved capacities (up to ~$NaC_{12}$) at 300 °C and 25 °C are comparable with the best values of hard carbon but the redistribution rate is faster by several orders of magnitude with effective diffusion coefficients as high as $10^{-4}$ $cm^2/s$. The storage is shown to fulfill the criteria of job-sharing storage with Na ions located on the alumina side of the hetero-contacts and electrons localized on the carbon side. The results are corroborated by electrochemical measurements (OCV, GITT, permeation) at room and high temperatures, by TEM and SEM microscopy, and chemical analysis. At high storage, a tendency to under potential deposition is concluded.

**Introduction**

Na-based room temperature batteries are the most promising alternative to lithium-based battery systems [1, 2, 3]. Its greatest asset is the very favorable abundancy (ranking number 4 if we ignore anion-forming elements). Apparent disadvantages concerning cell voltage and kinetics have been largely misjudged: The reversible cell voltage is not necessarily worse than Li, in fact the ionization potential of Na is even less than for Li. Lower voltage values for Na may be caused by higher lattice energies of analogous reaction products in the Li-case, but in praxi the reaction pathways are usually different. Furthermore, in contrast to half-cell potential, full cell potentials in liquid electrolytes are not influenced by the stronger solvation of Li. Unsurprisingly, the recent progress in Na based systems is remarkable [4, 5]. Yet, materials selection with respect to electrode and electrolyte is not yet optimal [1, 6]. This particularly concerns the anode (in the following we use this commonly used term instead of the more precise term 'negative electrode') where the rather costly, not very long-time stable and not very rapid "hard carbon" is used. It also affects SEI problems, in particular at the negative electrode side: in the case of liquid electrolytes, Na tends to give rise to the formation of SEI's that are frequently less stable than for comparable Li systems owing to the high reactivity and larger volume mismatch of Na systems [7, 8].

In this paper, we exploit the advantages of our job-sharing mechanism for Na storage [9]. Here, systems consisting of two-phases none of which exhibiting the possibility to store Na (i.e. $Na^+$ and $e^-$), may synergistically store Na. The requirement is that one phase stores the Na-ion and the other one the electron. Note that the relevance of the job-sharing storage was quantitatively explained by our lab and set out in various publications [9, 10, 11, 12, 13]. An independent clear support using magnetic measurements was recently published [14, 15].

The system Na-β˝-$Al_2O_3$ : graphite is predestined for this storage mode. A recent very systematic investigation [16] confirmed the general opinion that graphite bulk accommodates Na-ions only marginally but is of course able to quickly store the electrons, while contrarily Na-β˝-$Al_2O_3$ ($Na_2O·5Al_2O_3$) does hardly accommodate electrons (as we confirm in this paper, see Table S4-1, SI) but has room for further Na-ion storage in its conduction planes. In terms of Na-storability, the artificial electrode graphite : Na-β˝-$Al_2O_3$ is rather analogous to the silver storage in graphite:$RbAg4I_5$ composite showing - in contrast to the constituent phases - remarkable and very fast Ag storability (effective chemical diffusion coefficients as high as $4 \times 10^{-4}$ $cm^2/s$) [17]. It is worthy of note that – according to the present understanding - porous hard carbon mixed with liquid or solid electrolyte [18] is not a job-sharing electrode as the Na is supposed to be neutrally stored at the internal interfaces or pores of hard carbon whereby the electrolyte only provides the pathway for the ion transport to these sites. Actually, ongoing work in our lab offer an alternative view even for hard carbon.

**Results and Discussion**

In this way, composites of normal, soft graphite and the fast ion conductor Na-β''-$Al_2O_3$ can, as we show here, solve various problems. Owing to significant and ultrafast storage at low voltages (vs. Na), they have the potential to serve as very well working anodes the thickness of which could be much larger than for hard carbon. If Na-β''-$Al_2O_3$ is used as electrolyte (Spark plasma sintered Na-β''-$Al_2O_3$ shows fast Na conductivities at 300 °C (≥0.1 S $cm^{-1}$) and 0.003 S $cm^{-1}$ at room temperature (see Fig. S4-1)), no passivation layer is expected to form if Na β˝-$Al_2O_3$ is used as solid electrolyte (see SI part 3). In contrast to hard carbon, natural graphite does not need a sophisticated preparation procedure; moreover, alumina is highly abundant and it is only the synthesis of Na-β''-$Al_2O_3$ and the sintering procedure thereafter which require some efforts.

We will show that the storage mechanism fulfills the criteria of the job-sharing mechanism according to which Na diffuses heterogeneously (ion in alumina and electron in graphite) along the composite's heterointerfaces and is also stored heterogeneously, i.e. Na-ion at the ion conductor's side while the electrons are stored at the carbon side. A high interfacial area guarantees a significant storage capacity and the heterogeneous transport a very fast interfacial diffusion.

We first present our high temperature studies. As also for room-temperature graphite powder and Na-β˝- $Al_2O_3$ powder composites are ball-milled and hereafter subject to SPS sintering to obtain dense, phase-pure samples with negligible porosities (see Experimental Section for details). Additional Raman experiments confirm soft graphite (only $sp^2$) and exclude perceptible formation of $sp^3$ carbon (see Raman data, Fig. S4-3).

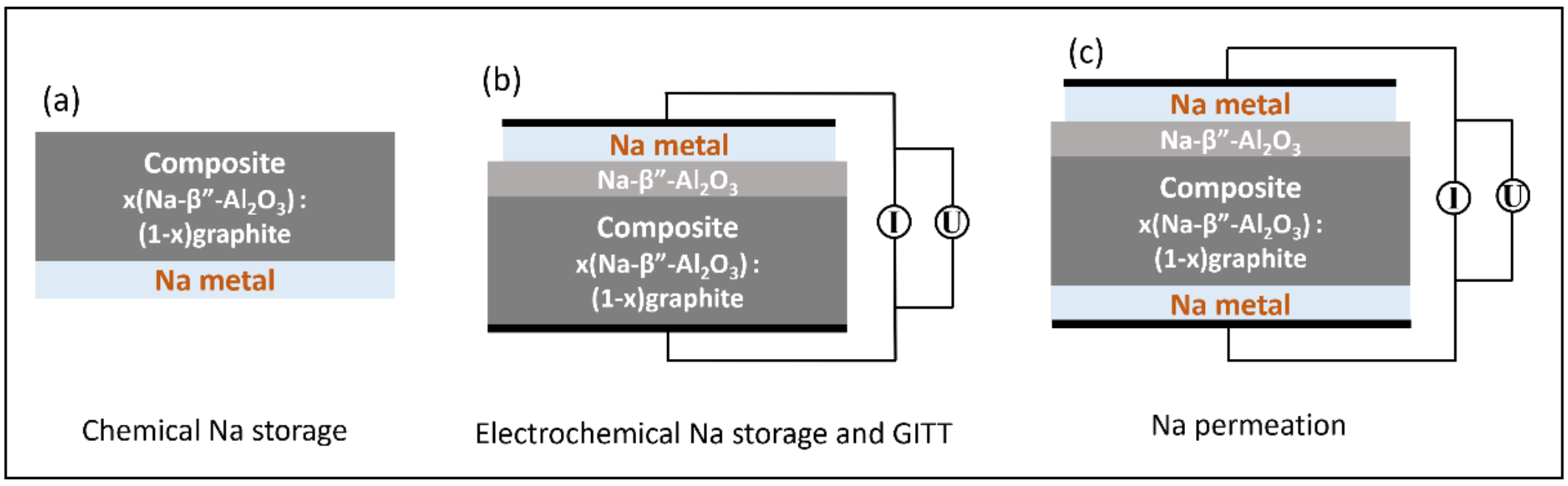

Fig. 1 Sketches of the set-ups for chemical Na storage (a), electrochemical Na storage and GITT (b), Na permeation (c).

The reason for first referring to the HT results is that at 300 °C we have precise experimental values for the thermodynamic limit of Na storage in the bulk of highly oriented graphite, from which we can conclude an upper bound of ~$NaC_{130}$ for the bulk storage in graphite [16]. The maximum Na storage in pure graphite powder at 300 °C has been shown to be $NaC_{50}$ (including Na adsorption on the surface of graphite powder, see Table S4-1). Fig. 1 sketches the set-ups for the various (electro)chemical methods we used. In the present contribution it is shown that Na storage in pure Na-β''-$Al_2O_3$ is minute (Fig. 2a). The observed blackening of the Na-β''-$Al_2O_3$'s surfaces on contact with Na is obviously due to color centers. According to our titration experiments, the molar fraction of Na storage in Na-β''-$Al_2O_3$ is less than $10^{-5}$ per lattice molecule (the precise value is below the experimental titration accuracy). In the composites, however, the excess Na content (corresponding to the storage) is found to be significantly higher. In fact, the observed values can be as high as ~$NaC_{12}$, comparable to hard carbon. These high excess values (Fig. 2a) must indicate interfacial storage. Naturally, the storage amount depends on the mixing ratio of the two phases constituting the composite. The maximum close to 50 % volume fraction is in agreement with interfacial effects and interfacial percolation. From the fact that graphite cannot accommodate Na or Na-ions to such an extent and Na-β''-$Al_2O_3$ can store Na-ions but not electrons to a measurable amount, we conclude that Na is indeed stored heterogeneously: Na-ions on the alumina side and electrons on the graphite side. Fig. 2b also shows that the total storage obtained electrochemically is compatible with the values obtained by chemical analysis (details of the titration method given in the experimental section).

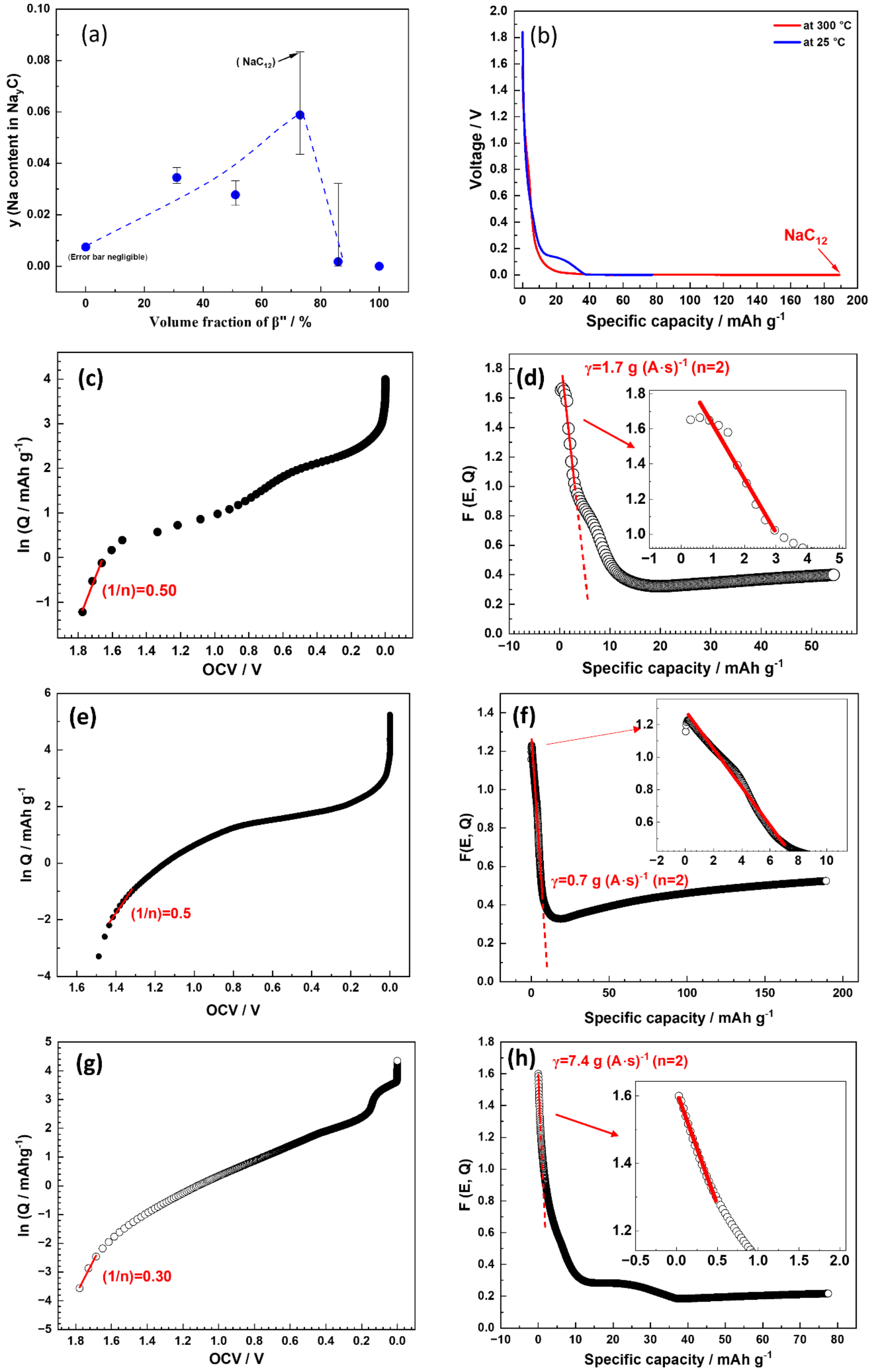

Fig. 2 (a) Na storage as a function of the volume fraction of Na-β″-$Al_2O_3$ (data obtained by chemical analyses after chemical Na storage at 300 °C). (b) Na storage as a function of OCV (specific capacities with respect to mass of graphite, obtained by charge accumulation of GITT measurements. 300 °C case was conducted at a current of 100 µA, and 25 °C case was conducted by 4 nm $TiO_2$ coated sample at a current of 5 µA). (c-h) Dependence of Na interfacial storage on OCV and hence on Na activity at 300 °C (c, d, e, f), and at 25 °C (g, h). For 300 °C case (c, d), the cell was cycled at a current of 5 mA before GITT measurement with an applied current of 1 mA; in (e, f), evaluation was directly conducted by GITT measurements with an applied current of 0.1 mA after relaxation into the current-less state. For the 25 °C case, sample was coated by 4 nm $TiO_2$, and the cell was cycled at a current of 100 µA before GITT measurement with an applied current of 5 µA. $F\ (E, Q) \equiv E\ + n(k_B T/e)\ \ln Q_m$.

Clear evidence of a job-sharing storage is provided by the thermodynamic analysis. In Refs.[9, 10] It was shown that for dilute situation (i.e. for low Na potential, high voltage) a power law for charge vs. voltage is expected to be fulfilled with an exponent $(1/n)$ between 0.5 and 0.25. An exponent of 0.5 is expected if we face two rigid layers (Na-ions in β″-$Al_2O_3$ and electrons in graphite), a slope of 0.25 is expected for two diffuse (Gouy Chapman) layers (0.33 if we face one rigid and one diffuse layer). At very small $Q$, intrinsic $Na^+$/$e^-$ concentrations increase $1/n$. At any rate, it is important to make sure to have filtered out additional (more sluggish) contributions. The latter point is the reason why the power law is practically well fulfilled for measurements under current (see Fig. S4-4). One should note, that neutral storage demands $1/n = 1$.

Figs. 2c, 2d show that power laws for GITT only result for high voltages (dilute range) with exponents of 0.5 (Note that the value at extreme OCV can be ascribed to intrinsic [12] and background effects). The realization of two rigid layers is compatible with the high intrinsic carrier concentrations. For higher storage a flattening is expected (as reflected by the smaller slope). Extending the validity range of our analysis by taking into account field effects (negligible for very low storage.), a much wider range can be fitted. For the samples shown in Fig. 2(d, f), realistic $\gamma$ parameters of 1.7 g $(A{\cdot}s)^{-1}$ and 0.7 g $(A{\cdot}s)^{-1}$ are obtained, respectively. As shown in the SI (part 2b), the $\gamma$ value depends on the microstructure and the temperature, and may hence scatter significantly. It is however important that in all cases, the $\gamma$ values are in approximate agreement with the measured capacities (Table S2-1), when applying the theoretical framework given in Ref. [10]. When applying a platelet model, we find, e.g. for the two 300 °C samples (Fig. 2d, 2f) that such $\gamma$ values are consistent with average plate thicknesses of 36 and 15 carbon layers sandwiched by Na-β″-$Al_2O_3$ layers, respectively (25 °C case is shown in Table S2-1). Note that these thicknesses of the graphite platelets would – in a monolayer model – be consistent with a Na content of $NaC_{18}$ and $NaC_8$, respectively. In view of the roughness of the model, these values are comparable with values derived by GITT ($NaC_{45}$ and $NaC_{12}$). The magnitude of the values is very approximate owing to the uncertainty in microstructure, charge separation and dielectric constant. At even higher storage the simple analysis has to fail owing to saturation and interactions. As shown in SI (part 1), the flattening of $F(Q)$ with $Q$ and the increase as very high $Q$ can be semi-quantitively explained by a partial and eventually full charge transfer corresponding to a tendency of neutralization (tendency to plating). In other words, at very high Na potentials (small OCV) transition to underpotential deposition to occur.

The microstructural model we have adopted in the above analysis, consists of percolating graphite-platelets of average thicknesses of less than 50 nm, contacted on both sides by Na-β˝-alumina, ideally mutually oriented parallel to the respective conduction planes. SEM, TEM (Fig. 3) and XRD (Fig. S4-5) show that this model is more or less realistic (more details, SI, part 1, 2). After ball-milling exfoliated graphite platelets are seen being mostly on the order of 50 nm or less, some are even graphene-like. After sintering we recognize alignment with Na β˝-alumina plates.

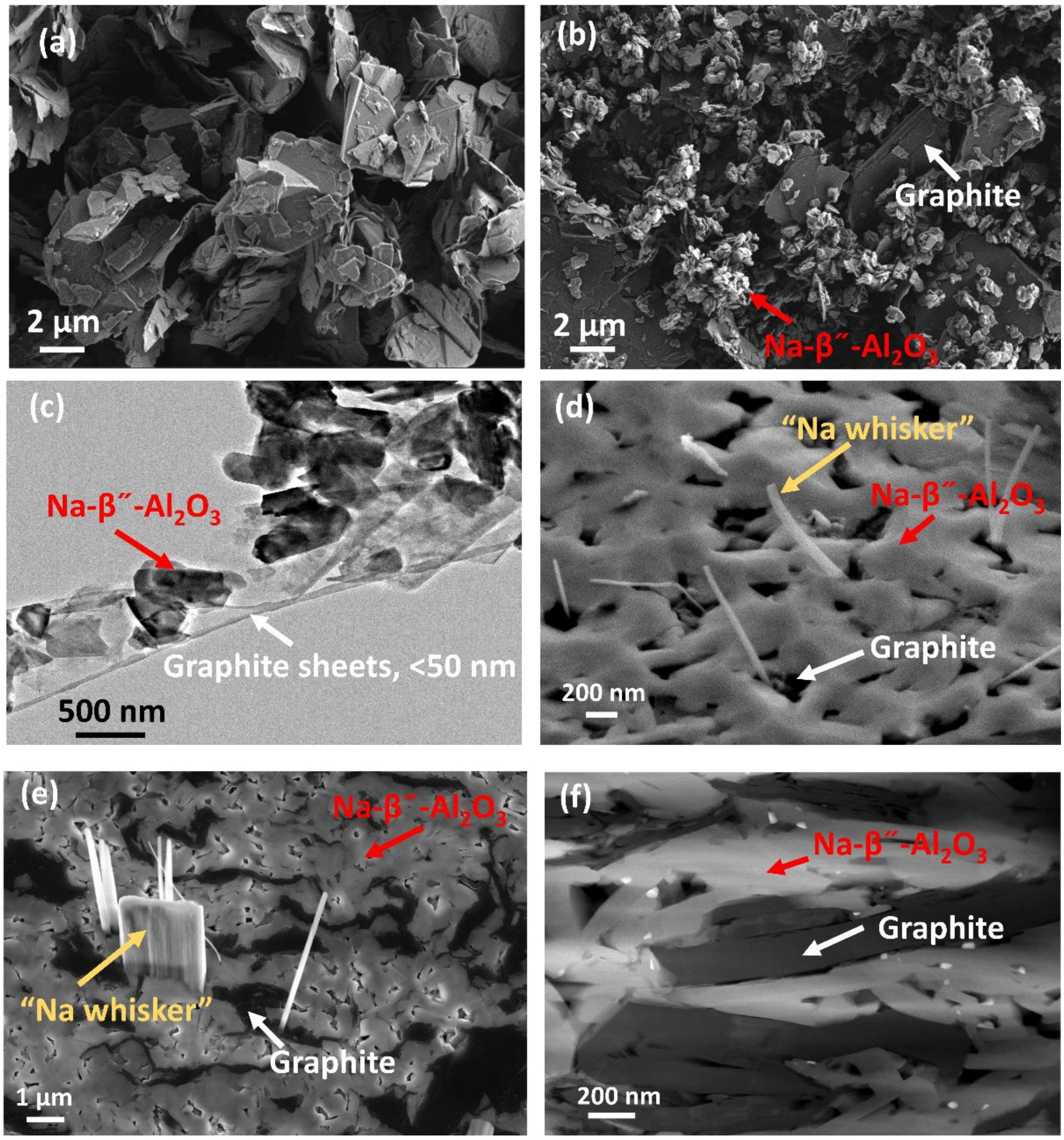

Fig. 3 (a) SEM image of original graphite. SEM image (b) and TEM image (c) of ball-milled mixtures. (d, e) SEM image for the composite pellet after contact with Na metal at 25 °C for 2 minutes. The emerging "Na whisker" may be $Na_2O$ or $Na_2CO_3$. (f) TEM image of composite pellet after chemical Na storage at 300 °C for 14 days.

The kinetic analysis (Fig.4a, c, d) from GITT measurements shows an extraordinarily good agreement with the prediction of an exponential long-time behaviour. The analysis reveals extremely high effective diffusion coefficients of $10^{-4}$ $cm^2/s$ (For details, see ref. [17, 19]). One should note that comparable values for hard carbon are about 4-6 orders of magnitude lower [20, 21]. Fig. 4b gives the $D_{eff}$ values as a function of OCV (i.e. concentration). The explanation of the details is beyond the scope of this paper.

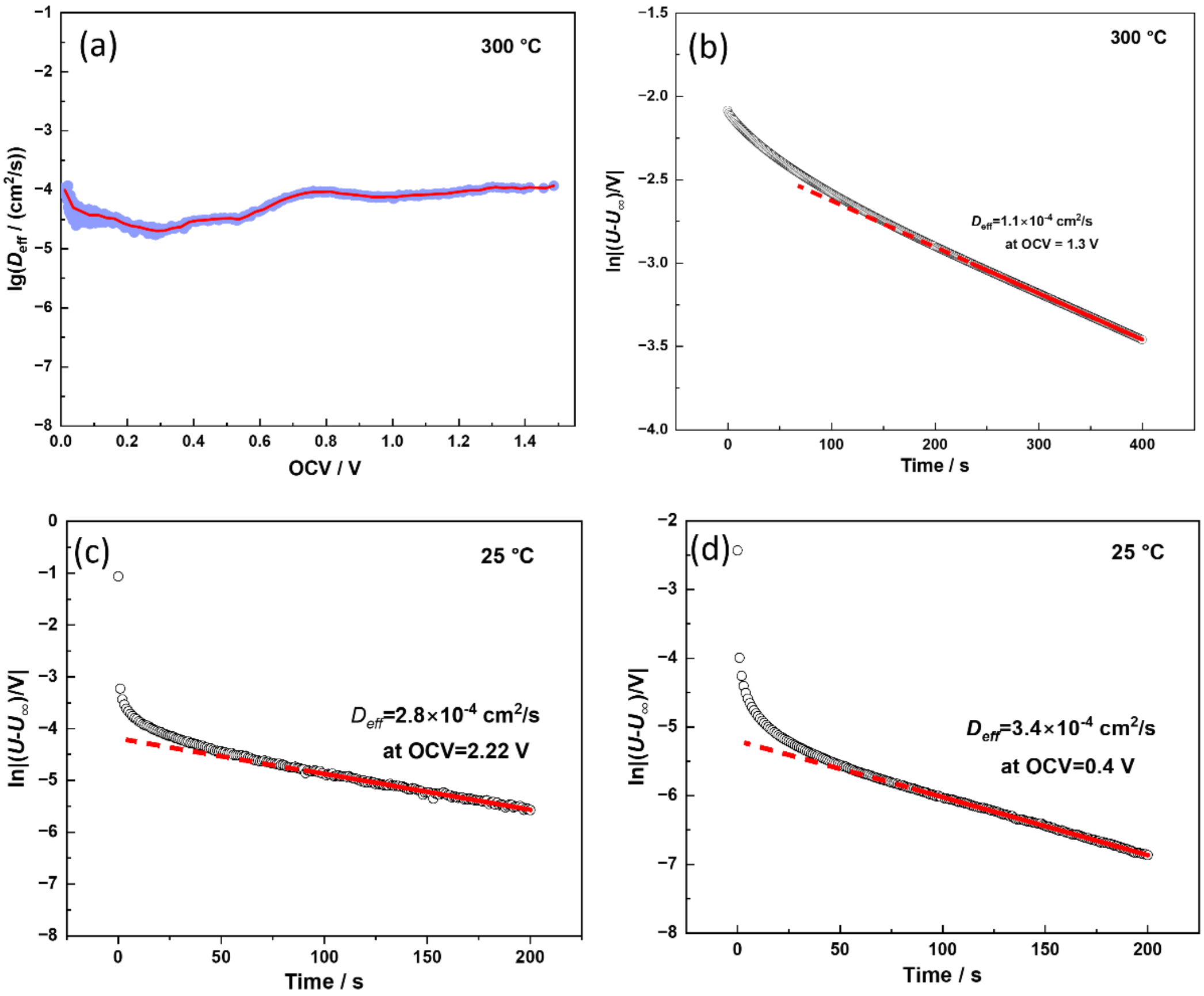

Fig. 4 $D_{eff}$ values as a function of OCV (i.e. concentration) (a). Evaluation of diffusion coefficients for the long-time relaxation at 300 °C (a) and at 25 °C (c, d) from the linear parts of the curves (Coefficient of determination of the fittings, $R^2$>99.99%).

In $Rb_4AgI_5$ : graphite [17] with its monolayer morphology we had found similarly high values for $D_{eff}$. There we could ignore super-capacitive contribution. In our case, this may apply as well, as the ionic transport perpendicular to the conduction planes is poor [19, 22].

We also set-up a permeation experiment to check the speed of Na transport (Fig. 1c). After reaching electrochemical equilibrium by Na having diffused through the primarily electronically composite, a zero value for the voltage is to be attained (See Fig. S4-6). It turns out that the rates of "chemically charging" the cell (applying Na metal) as well as the rate of thermalization in cases where the temperature was varied, are lower than the permeation speed itself leading only to a lower bound for $D_{eff}$ which is however compatible with the GITT measurements.

Strikingly, SEM investigations of the permeation studies display two-dimensional extrusions of oxidized Na ($Na_2O$ or $Na_2CO_3$) coming out from the interfaces between the two phases (Fig. 3d, e). (Here in a purely chemical experiment, we had attached Na to the adjacent side of the composite and observed the quick Na formation on the initially clean surface at the interfaces between carbon and β''-$Al_2O_3$) confirming interfacial transport.

Now having established the mechanism, we turn to the room temperature measurements. They essentially show the same favorable features as obtained at 300 °C.

Fig. 2b shows that similarly as at high temperature the overall capacity is high (up to $NaC_{12}$) and the distribution speed in the composites is enormous. Again the evaluation of the effective diffusion coefficient succeeds very well delivering $D_{eff}$ ~$10^{-4}$ $cm^2 \cdot s^{-1}$. The more detailed dependence on voltage (composition) is given in SI. As for the HT samples, permeation results (Fig. S4-6) give a consistent upper limit compatible with the GITT experiments. The agreement with chemical analysis is shown in Fig 2a. Because of interfacial limitation we cannot obtain the equilibrium storage; yet from the results we can give a lower bound for the diffusion coefficient which is compatible with the high $D_{eff}$ -values found from GITT. The fast chemical storage of Na in the composite confirms the potential aptitude as Na battery anode.

As at higher temperatures the thermodynamic analysis fully conforms with a job-sharing mechanism. The analytical relationships for the evaluation of OCV as a function of $Q$ succeeds equally well with a power of 0.5 and $\gamma$ values comparable to 300 °C (note that $\gamma \propto \frac{1}{T}$ ), leading to the same conclusions as above: A percolation of graphite platelets of less than 50 nm thickness on the average contacted on both sides by oriented Na-$\beta''$-$Al_2O_3$ fully explains the total charge as well as its dependence on OCV.

Preliminary results on the full cell suffer from not having optimized internal cell resistances. At high temperatures where we use Na-$\beta''$-$Al_2O_3$ as electrolyte, the cell resistance is governed by the typical solid-solid contact resistances which let the cell completely fail after 10 cycles at atmosphere pressure (see Fig. S4-7). Here application of pressure (as usual for solid electrolyte cells) will be helpful. At low temperatures, the conductivity of the ceramic electrolyte is rather poor owing to grain boundaries as impedance spectroscopy reveals. Liquid electrolytes exhibit the typical problems with passivation layers. Finding the appropriate liquid electrolyte has also to face the enhanced basicity of the electrode due to the $Na_2O$ content of Na-$\beta''$-$Al_2O_3$. $TiO_2$ films [16] as artificial passivation layer is proved helpful also in terms of precluding co-intercalation, but is also characterized by a significant resistance (Fig. S4-8).

Optimization of full cells definitely exceeds the scope of this paper. The effect of the internal resistance is that complete charge or discharge takes substantial time. The low currents however suffice for revealing the unprecedented properties of the synergistic electrode in terms of storage and Na-transport within.

**Summary**

The present paper has various highlights. It is a further example of the job-sharing storage mode which has been promoted and described in previous publications. In view of the crudeness of the model and complexity of the microstructure, the agreement with the storage capacity and the thermodynamic analysis is remarkable. It is the first example of a practically useful electrode based on this principle. The composite used based on Na-β''-$Al_2O_3$ and graphite is not only an inexpensive alternative to hard carbon as anode in Na-batteries, it shows not only comparable capacities, but shows unprecedented high diffusion rates due to the enormously high heterogeneous diffusion coefficient. As a consequence, a much higher mass could be deployed. Absence of the typical anode-electrolyte interface problems when Na-β''-$Al_2O_3$ is used as electrolyte which are particularly severe for Na is a further asset.

## Experimental

### Materials

Graphite powder (2~3 μm) was supplied by NGS Naturgraphit GmbH. Na-β''-$Al_2O_3$ powder (1 μm) was purchased from MSE Supplies LLC. Sodium metal was purchased from Sigma-Aldrich (99.9%, cubes in mineral oil).

### Scanning electron microscopy (SEM)

SEM images of the graphite, Na-β''-$Al_2O_3$ powders, and composite of Na-β''-$Al_2O_3$ and graphite were taken using a Zeiss Merlin field emission scanning electron microscope equipped with a Gemini II column.

### Materials mixture by Ball milling

The ball milling was carried out by planetary ball milling equipment with milling beads $ZrO_2$ (pulverisette 5, Fritsch GmbH). Normally 3~4 g powders of Na-β''-$Al_2O_3$ and graphite at given volume ratios (e.g. 72% Na-β''-$Al_2O_3$ to 23 % graphite). To mix the powders, 6~10 ml 2-propanol was added as milling solvent. The milling was carried out at 100 rpm for ~10h. After that, to remove solvent, the mixtures were dried at room temperature in a fume hood for ~8 h, then the mixtures were transferred to vacuum dry at 80 °C for more than 12h.

### Spark plasma sintering (SPS)

SPS was carried out using a Type HP D direct current pulses furnace (FCT Systeme GmbH). All pellet samples were fabricated with a 10 mm graphite die and corresponding graphite punches. To sinter pellet composites, 0.5~0.7 g mixture and 0.3 g Na-β''-$Al_2O_3$ were used. For each sintering process, the sample was packed and loaded at 6 KN, and the chamber was vacuumed and filled with Ar at a pressure of 4000 Pa. Then the sample was heated to 1100 °C at a rate of 100 °C/min, then to 1200 °C at a rate of 50 °C/min, and was held at 1200 °C for 5 min before it was allowed to cooled naturally to room temperature. The as-obtained pellet was hand polished with sandpapers before subsequent characterization/experiment.

### Na storage experiments and electrochemical measurements

All Na storage experiments were performed using the homemade high temperature set-up as depicted in Fig. 1 and Fig. S4-1. For the electrochemical Na storage performed on the composite pellet containing 80 wt. % Na-β″-$Al_2O_3$, a piece of Na was placed above the pellet, and the cell was assembled and sealed inside the glovebox. Then the cell was taken out from the glove box, connected to a Source Meter electrochemical workstation and heated in a vertically placed furnace, during which open circuit voltage of the cell was measured. After reaching 300 °C, galvanostatic discharge and constant voltage discharge measurements were performed accordingly. Details for AC impedance measurements are shown in ref. 16.

For the chemical Na storage performed on the composite pellets, the piece of Na was placed below the pellet and the cell was not connected to the electrochemical workstation. After Na storage, extra Na metal at the bottom of the pellet was removed before subsequent characterization.

**Chemical analysis**

The contents of chemically Na storage inside the composites were determined by titration. Typically, ~20 mg of the sample was placed in a Schlenk flask with 80 ml of freshly prepared bi-distilled water, and heated under Ar atmosphere for approximately 20 h at 90 °C. Under these conditions the intercalated Na reacts with $H_2O$ according to $Na + H_2O \rightarrow Na^+ + OH^- + 0.5H_2$. Still under Ar atmosphere, the formed $OH^-$ was titrated with HCl using a Metrohm Titrino 877 automatic titrator and the amount of $OH^-$ corresponds directly to that of the intercalated Na. There is also a certain amount of $Na_2O$ which is leached out from the pure Na-β″-$Al_2O_3$. This blank value was subtracted. It was made sure that this blank value did not depend on Na-β″-$Al_2O_3$ particle size.

**X-ray diffraction (XRD)**

XRD data were collected on a Malvern Panalytical Empyrean X-ray diffractometer using Cu-Kα radiation operating at 40 KV and 40 mA. All the Na stored samples were transferred to a polycarbonate domed sample holder inside the glove box before characterization.

**Raman spectroscopy**

Raman spectra were acquired in a backscattering configuration using a home-built micro-Raman spectrometer with a diode laser (excitation line 532 nm) and a Nikon CF IC EPI Plan SLWD ultra-long working distance objective (X50, 13.8 mm working distance, numerical aperture 0.45). All the Na stored samples were transferred to a sealed sample holder with a glass window inside the glove box before characterization.

**Conductivity measurements**

The electronic conductivities were measured using the standard van der Pauw technique with a Keithley 2400 Source Meter at room temperature. For measuring the ionic conductivity, the pellets were first annealed in the air at 800 °C for 10 h to remove graphite and deposited on both sides with 400 nm Pt by DC sputtering in the glovebox. Then the pellets were transferred to an airtight test cell capable of allowing Ar to flow. The ionic conductivities were derived from their electrochemical impedance spectra taken by a Novocontrol Alpha-A analyzer at 300 °C.

**Acknowledgements**

The authors are grateful for R. Merkle (chemical analysis), F. Kaiser (set-up design), A. Sorg (SPS, XRD), V. Duppel (TEM and SEM), Dr. H. Wang and Prof. P. van Aken (TEM), B. Fenk (SEM), A. Fuchs (SEM), H. Hoier (XRD), A. Schulz (Raman), Prof. E. Kotomin (internal review), and U. Traub, Dr. J. Popovic-Neuber, U. Klock, Dr. Y. Wang for their technical support. Y. Zhu acknowledges the support from the National Natural Science Foundation of China (No.22475201) and the Shangdong Provincial Natural Science Fund for Excellent Young Scientists Fund Program (Overseas) (2024HWYQ-036). Support from MPG is gratefully acknowledged.

## Supporting Information

### 1. Explanation of $F(Q)$ and $ln\ Q(E)$ for high activity

#### a) Activity coefficients

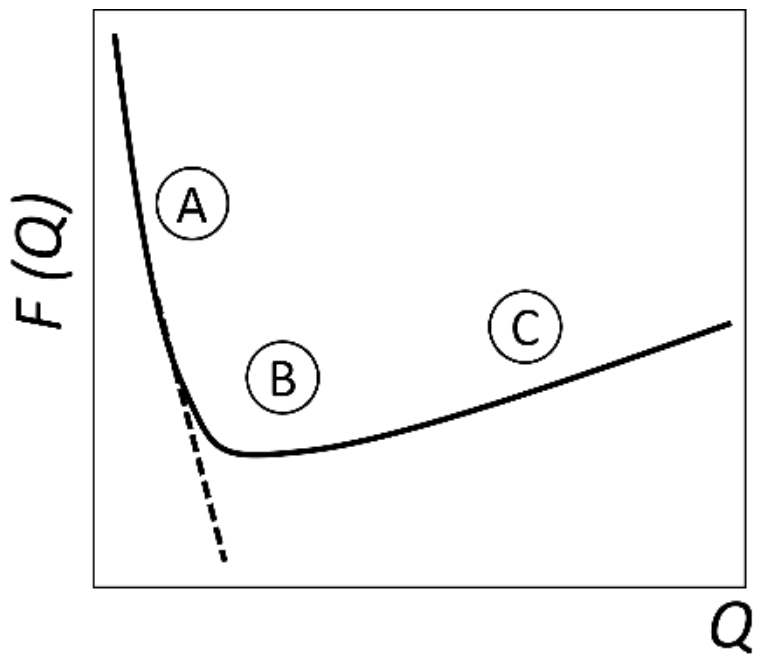

Fig. S1-1 Sketch of $F(Q)$ with $Q$

$$F(Q) \equiv E + \alpha n \ln Q \tag{1}$$

where $E$ is voltage (OCV), and $\alpha = \frac{k_B T}{e}$.

With activity coefficients

$$F(Q) \equiv E + \alpha n \ln Q = -\alpha\gamma Q + constant - \alpha n \ln \bar{f} \tag{2}$$

Simplification:

n=2, two rigid layers

$$F(Q) \equiv E + 2\alpha \ln Q = -\alpha\gamma Q + constant - 2\alpha \ln \bar{f}$$

$\bar{f} = \sqrt{f_{Na_i^{\cdot}(\beta)} f_{e'(C)}} < 1$, meaning attraction. (3)

Hence the slope becomes flatter (Ⓑ, Fig. S1) and even inverse (Ⓒ, Fig. S1), for high $Q$, $\bar{f} < 1$, attraction.

#### b) Mechanism

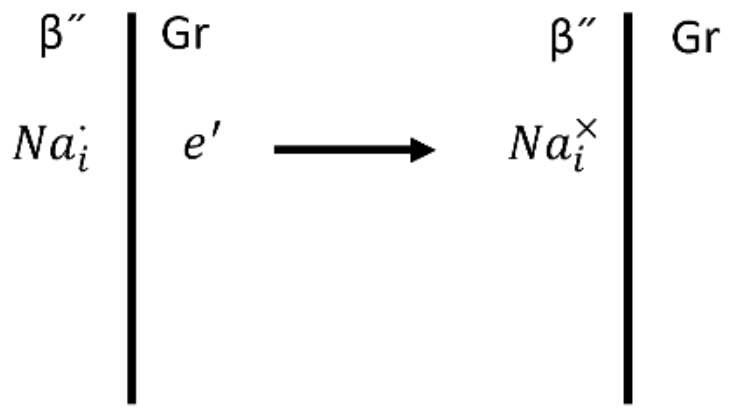

Fig. S1-2 Attractive interaction

Semi-qualitatively:

$$F(Q) \equiv E + \alpha n \ln Q = -\alpha\gamma(Q - \mathrm{F}[Na^{\times}])$$

$$= -\alpha\gamma Q + \mathrm{F}\alpha\gamma[Na^{\times}] \qquad (4)$$

Note that $Q$ and $[Na^{\times}]$ are both normalized to the mass of graphite. F is Faraday constant.

More precisely $n$ is a function of $Q$ ($n \to 1$)

$$F(Q) \equiv E + \alpha n \ln Q = -\alpha\gamma(Q - \mathrm{F}[Na^{\times}]) = -\alpha\gamma Q + \mathrm{F}[Na^{\times}] + \alpha \ln Q \qquad (5)$$

both latter terms lead to flattening and inversion.

More quantitatively:

$$\mathrm{Na} \rightleftharpoons \mathrm{Na_i^{\cdot}} + \mathrm{e_C'} \qquad [Na_i^{\cdot}][e_C'] = a_{Na}K\varkappa \qquad (6)$$

$$\mathrm{Na} \rightleftharpoons \mathrm{Na_i^{\times}} \qquad [Na_i^{\times}] = a_{Na}K' \qquad (7)$$

$$\mathrm{Na_i^{\cdot}} + \mathrm{e_C'} \rightleftharpoons \mathrm{Na_i^{\times}} \qquad [Na_i^{\times}] = [Na_i^{\cdot}][e_C']\frac{K'}{K\varkappa} \qquad (8)$$

$$\frac{[Na_i^{\times}]}{[Na_i^{\cdot}]^2} = \frac{K'}{K\varkappa} \quad \Rightarrow \quad \frac{[Na_i^{\times}]}{[Na_i^{\cdot}]} = \sqrt{\frac{K'}{K\varkappa}}\sqrt{[Na_i^{\times}]}$$

$$= \frac{K'}{\sqrt{K\varkappa}} \cdot a_{Na}^{1/2} \qquad (9)$$

$\frac{[Neutral]}{[ionized]} \propto a_{Na}^{1/2}$ increases with increasing $a_{Na}$!!

**c) Explanation of $F(Q)$ curve by ongoing neutralization.**

This will be evaluated in a following paper.

**2. Interfacial storage by job-sharing mechanism**

Based on TEM images of ball-milled composites, it is pertinent to state that after synthesis the composites consist of nano-sized graphite sheets (<50 nm, even single layer graphene) sandwiched by Na-β″-$Al_2O_3$ (Fig. S2-1). If all the platelets are well connected, if suffices to use a single element.

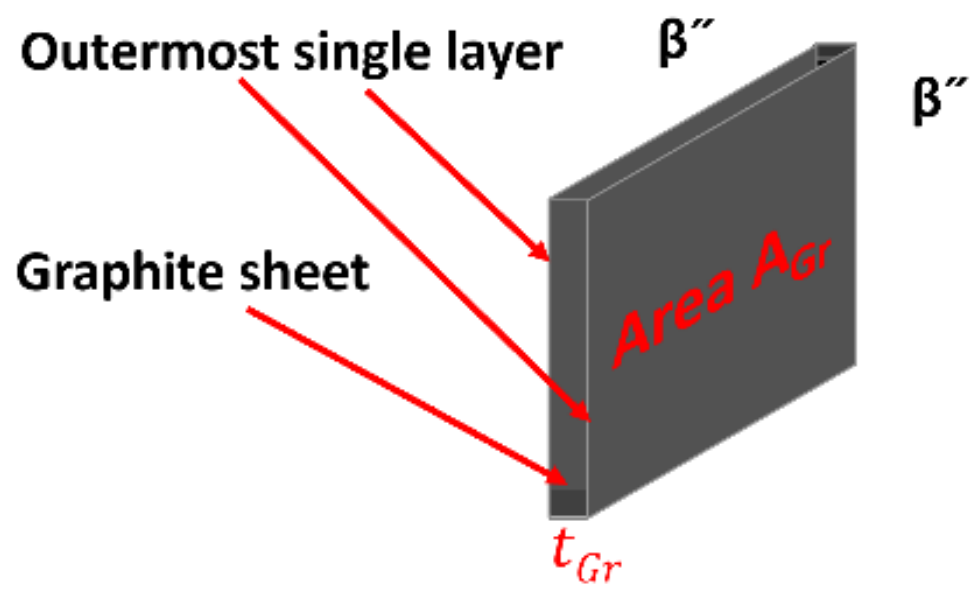

Fig. S2-1 Sketch of graphite sheet covered by Na-β˝-$Al_2O_3$ (β˝). Ina rigid storage model, only the outermost single layers are charged (by electrons)

**a) $n$, $F(E,Q)$ and $\gamma$ obtained by experimental data**

According to Ref. 10 and Fig. S2-1, we have

$$exp^{-\frac{eE}{k_BT}} \propto a_{Na} \propto Q^n exp^{\gamma Q} \tag{10}$$

where $e$ is elementary charge, $E$ voltage (OCV), $Q$ charge capacity, $a_{Na}$ activity, $k_B$ Boltzmann constant, and $T$ temperature.

$n$ determined by $exp^{-\frac{eE}{k_BT}} \propto Q^n$ i.e. by the small $Q$ approximation ($exp^{\gamma Q} \approx 1$).

$$F(E,Q) = E + n\frac{k_BT}{e}\ln Q \tag{11}$$

If we plot $F(E,Q)$ as a function of $Q$, we should obtain a line with $\frac{-\gamma}{k_BT/e}$ as slope.

**b) Theoretical $\gamma$ calculation based on job-sharing mechanism**

According to ref. 10, we have

$$\gamma = \frac{\mathrm{F}s}{\epsilon\epsilon_0 RT}\rho_{Gr}\frac{V_{Gr}}{A_{Gr}} \tag{12}$$

As we refer the charge to the mass of graphite. $s$ is distance between the adjacent layers of the graphite and Na-β˝-$Al_2O_3$. $\rho_{Gr}$ is the density of graphite sheet, $V_{Gr}$ the volume of graphite sheet, $A_{Gr}$ the surface area of the large planes of the platelets. $\epsilon$ is taken as 10. Using $t_{Gr}$ as thickness of graphite sheets ($t_{Gr} = \frac{V_{Gr}}{A_{Gr}}$)

we have

$$t_{Gr} = \frac{\gamma\epsilon\epsilon_0 RT}{Fs\rho_{Gr}} \tag{13}$$

Substituting $\gamma$ values derived from experimental data (GITT), the thicknesses of graphite sheets can be compared with the thicknesses of graphite sheets derived by GITT measurements (see Table S2-1).

Table S2-1 Thickness of graphite sheets (carbon layers) calculations

| Sample | $\gamma$/ g $(A.s)^{-1}$ derived from GITT | Thickness derived from $\gamma$ / nm | Composition derived from $\gamma$ | Thickness derived from GITT / nm | Composition derived from GITT | Capacity derived from GITT/ (A.s) $g^{-1}$ |
|---|---|---|---|---|---|---|
| S1 300 °C | 1.7 | 12 (36 layers) | $NaC_{18}$ | 30 (90 layers) | $NaC_{45}$ | 14 |
| S2 25 °C | 7.4 | 24 (70 layers) | $NaC_{35}$ | 19 (56 layers) | $NaC_{28}$ | 22 |
| S3 300 °C | 0.7 | 5 (15 layers) | $NaC_{8}$ | 8 (24 layers) | $NaC_{12}$ | 53 |

### 3. Stability of contact of Na-β″-$Al_2O_3$ with Na

$$6\mathrm{Na} + 2\mathrm{Al_2O_3} \rightleftharpoons 3\mathrm{Na_2O} + 4\mathrm{Al}$$

has a positive $\Delta_R G$ ①, but it may be slightly negative ②if

$$6\mathrm{Na} + (3\mathrm{x}+1)\mathrm{Al_2O_3} \rightleftharpoons 3\mathrm{Na_2O}\cdot\mathrm{xAl_2O_3} + 2\mathrm{Al}$$

i.e. $Na_2O$ is dissolved with Na-β″-$Al_2O_3$ phase. Yet an SEI phase was never reported.

Note: ① Thomas B. Reed, Free Energy of Formation of Binary Compounds, MIT Press, Cambridge, MA, (1971).
② M. Barsoum, Critical review of the thermodynamics of the Na-Al-O system, 25 4393–4400 (1990).

### 4. Supporting Figures and Tables

Table S4-1 Average Na concentrations at 300 °C by chemical analysis

| Sample | Pure HOPG | Pure graphite powder | 40 wt.% β″-$Al_2O_3$ | 60 wt.% β″-$Al_2O_3$ | 80 wt.% β″-$Al_2O_3$ | 90 wt.% β″-$Al_2O_3$ | Pure β″-$Al_2O_3$ |
|---|---|---|---|---|---|---|---|
| Composition | $NaC_{130}$ see Ref. 16 | $NaC_{50}$ | $NaC_{27}$ | $NaC_{36}$ | $NaC_{17}$ | $NaC_{558}$ | below test limit (<$10^{-5}$ per molecule) |

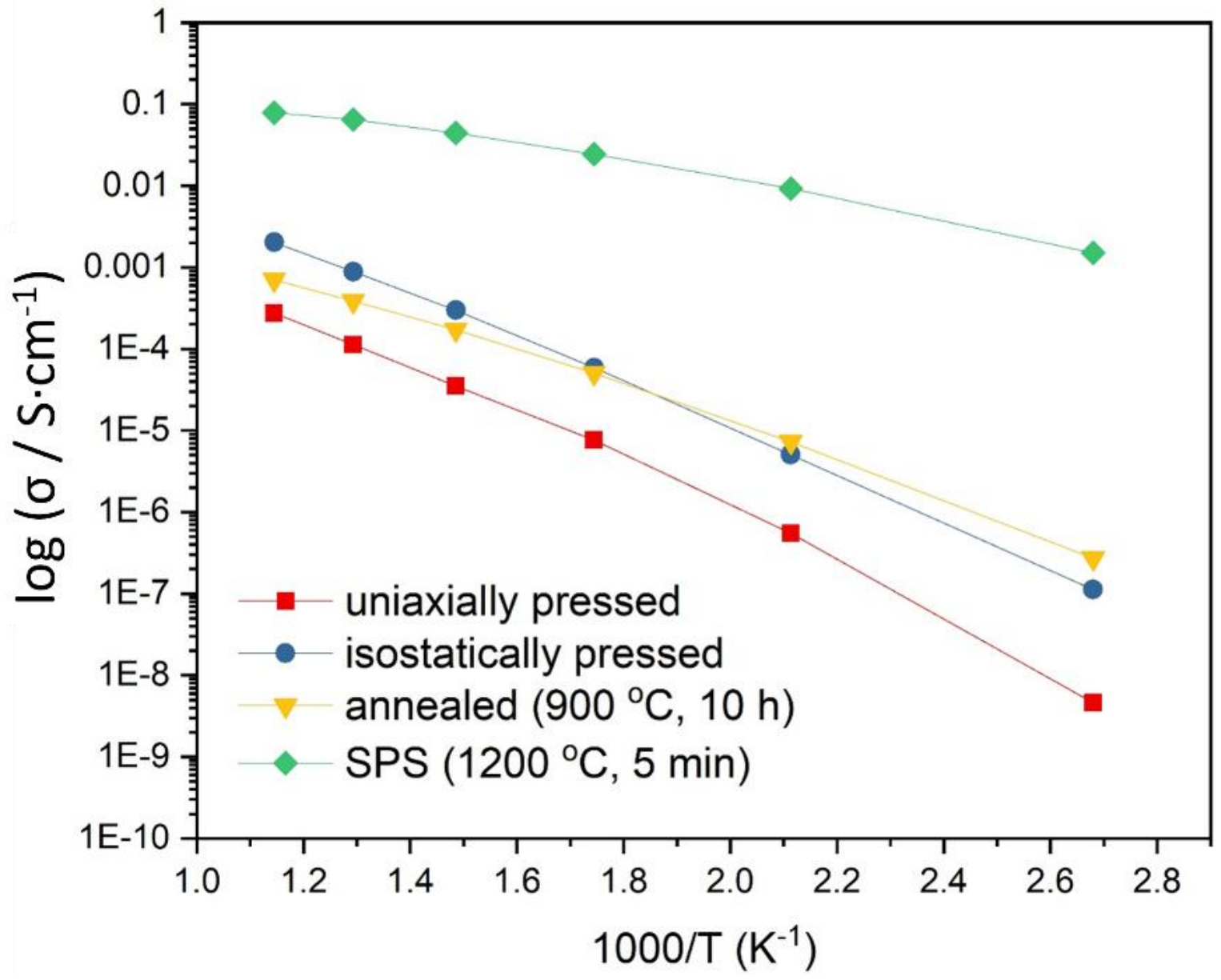

Fig. S4-1 Ionic conductivity of Na-β˝-$Al_2O_3$ ceramics. The values at high temperatures for SPS-synthesized samples are limited by the wiring and only give us lower limits.

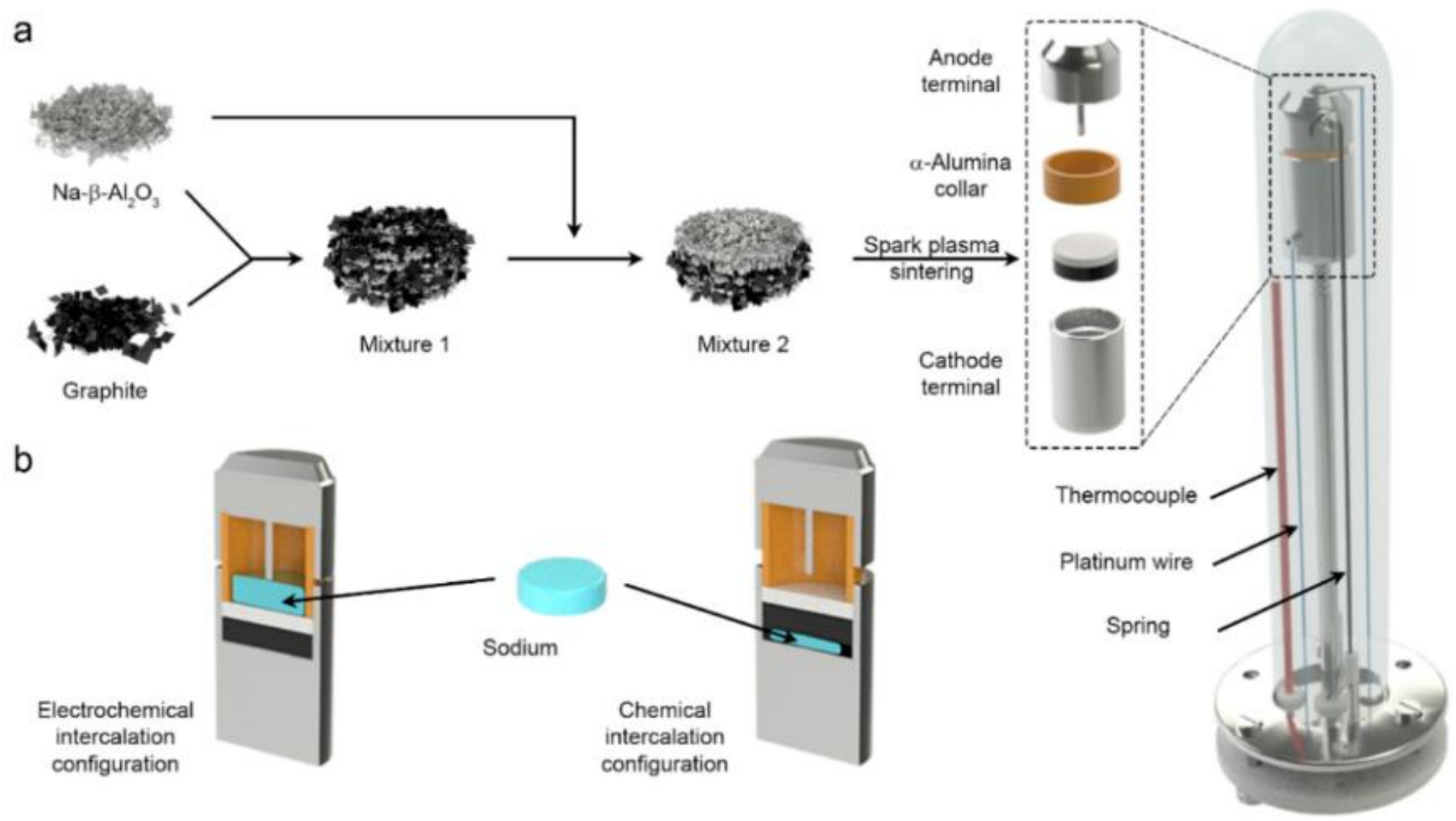

Fig. S4-2 Schematic illustration of a) the fabrication process for a Na-β˝-$Al_2O_3$/graphite pellet and the 300 °C two-electrode cell design, and b) the configurations for (electro)chemical Na storage.

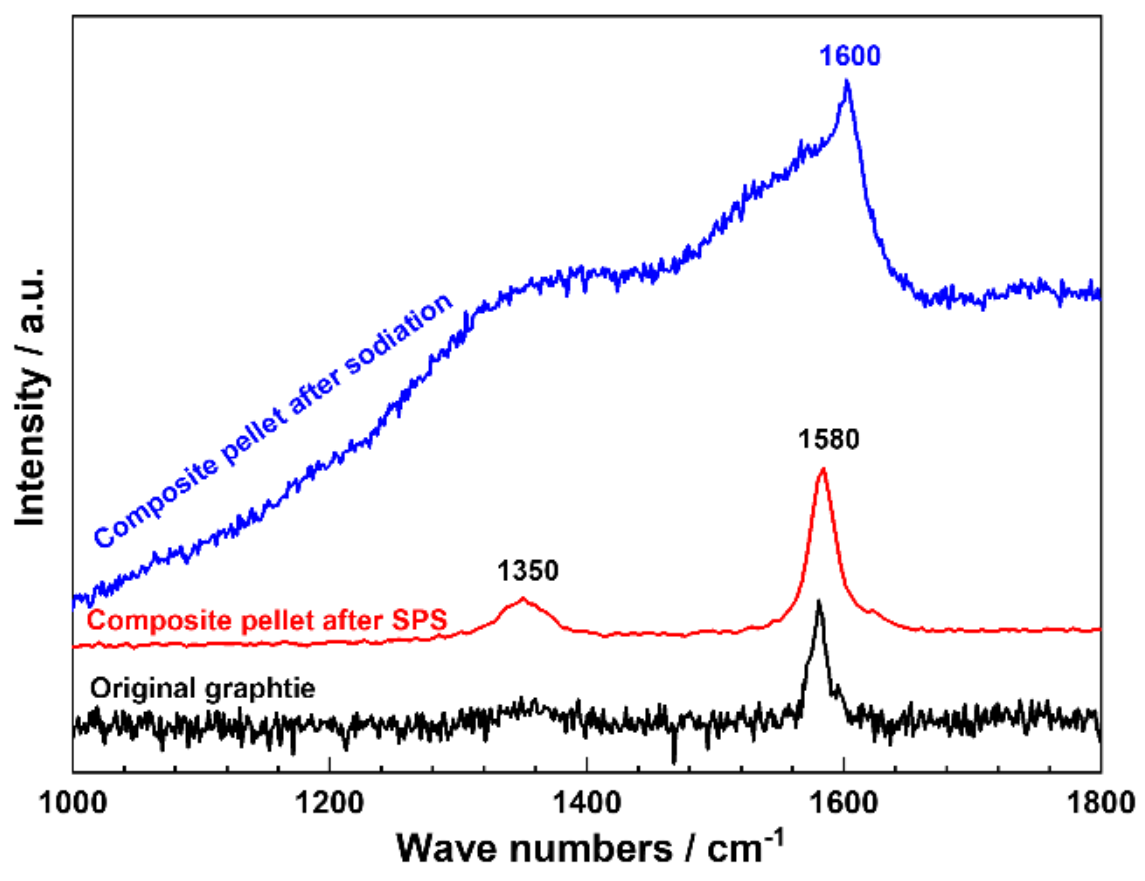

Fig. S4-3 The Raman spectrum shows that $sp^2$ bonds (1580 $cm^{-1}$) dominate. The blue curve reveals a strongly broad Na-peak which is due to not having removed all Na metal (chemical sodiation by direct contact with Na at 300 °C for 14 days. Composite: 80 wt. % Na-β″-$Al_2O_3$-20 wt. % graphite).

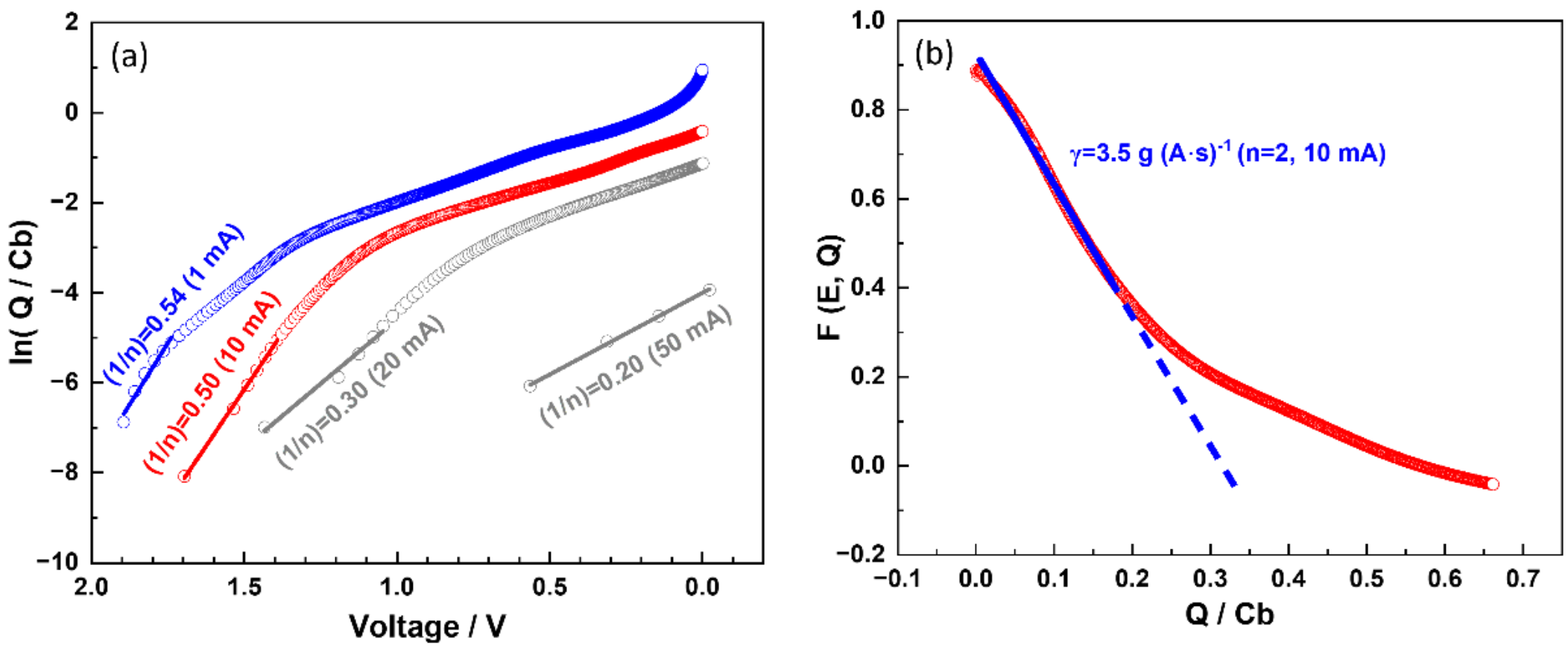

Fig. S4-4 Dependence of Na interfacial storage on Na activity by Galvanostatic discharge at 300 °C. The fact that power laws here are well fulfilled is due to the fact that under current other contributions to the OCV are suppressed. Naturally, current dependencies of the overvoltage need to be small in order to neglect the influence of current dependent over-voltages (composite: 80 wt. % Na-β″-$Al_2O_3$-20 wt. % graphite).

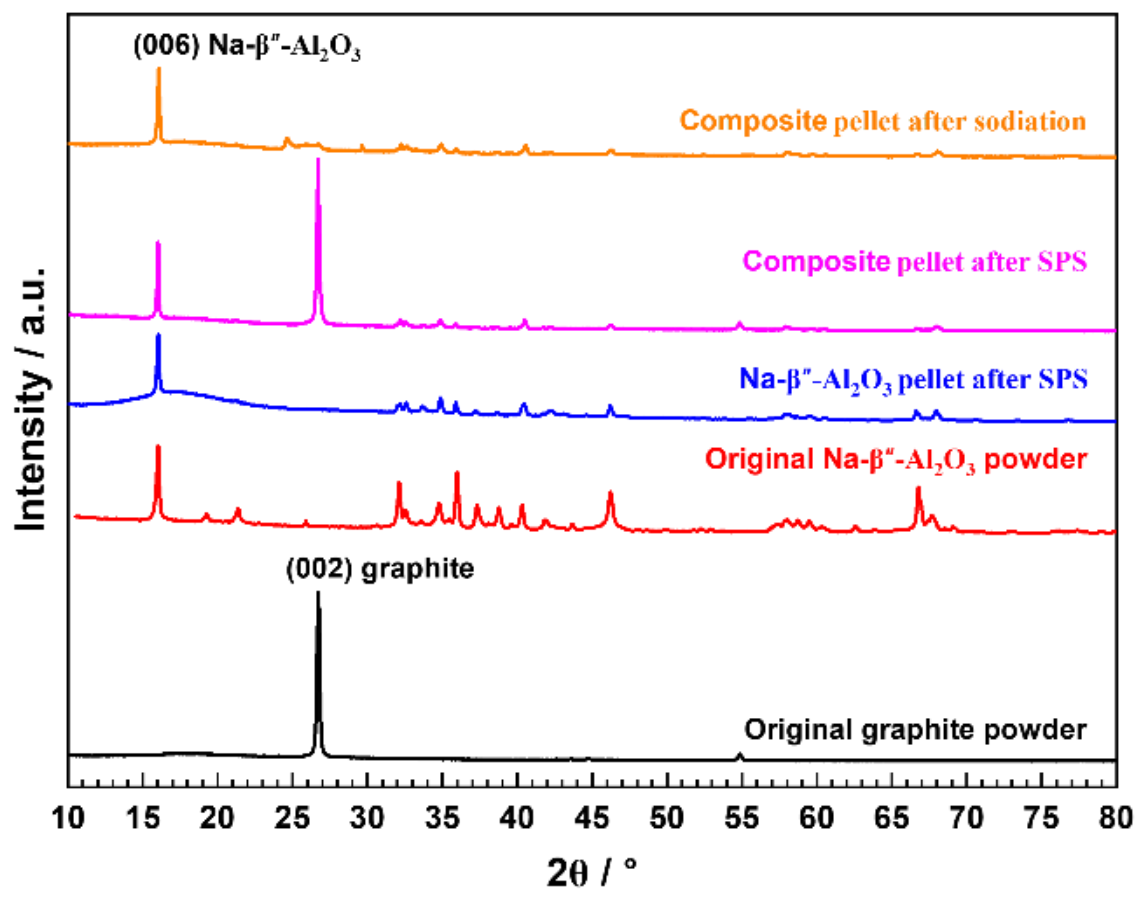

Fig. S4-5 XRD patterns (The composites refer to 80 wt. % Na-β″-$Al_2O_3$-20 wt. % graphite. Composite sodiation was conducted at 300 °C for 14 days. Note that (002) peak after sodiation shifts to left and is weak).

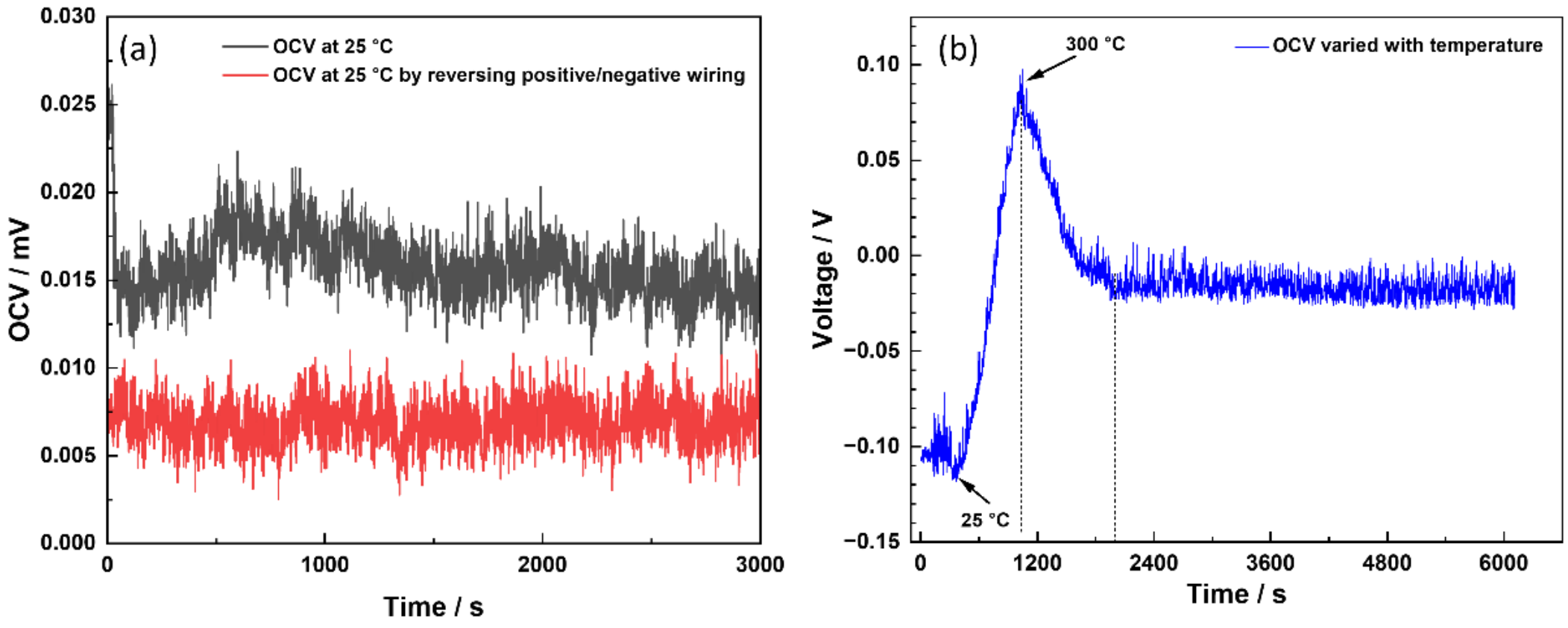

Fig. S4-6 OCV at 25 °C (a) and at dynamical temperatures (b). Data obtained by Na permeation experiment. Cell assembly cost about 20 min before OCV measurement.

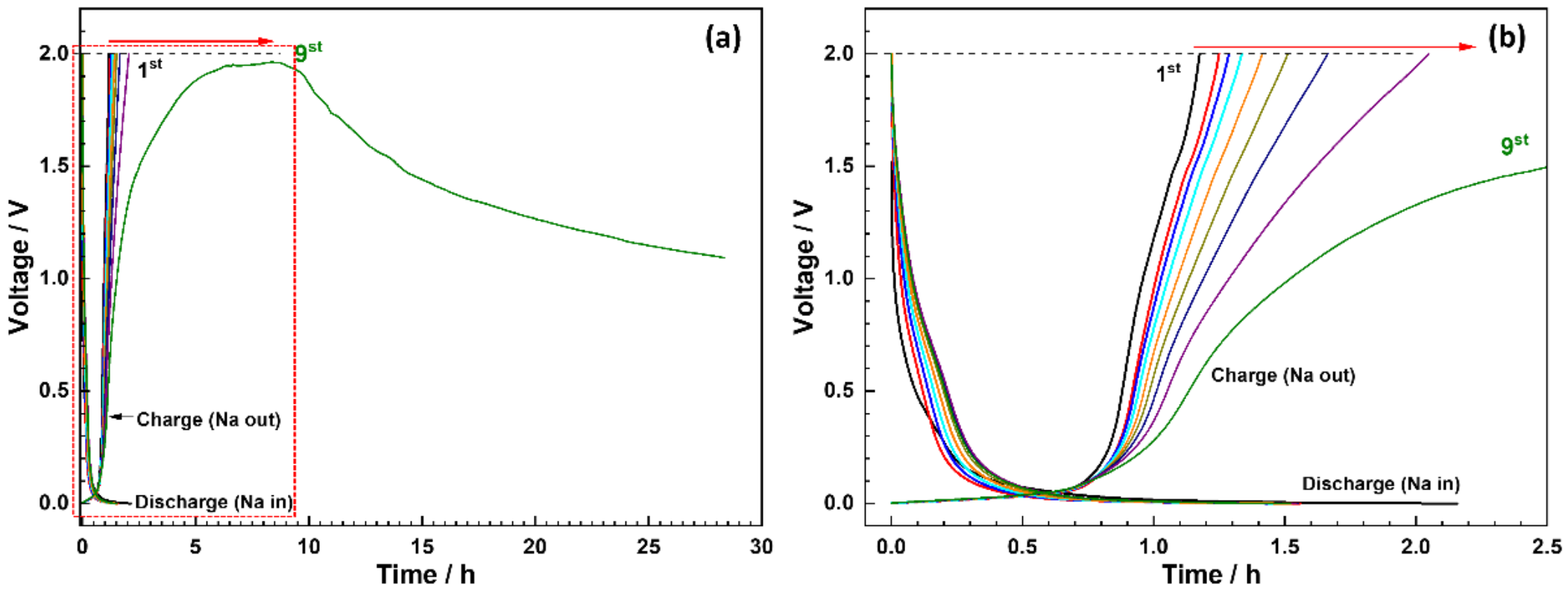

Fig. S4-7 Galvanostatic discharge/charge for cell cycled at a current of 100 µA at 300 °C, at atmospheric pressure (composite 80 wt. % Na-β˝-$Al_2O_3$-20 wt. % graphite).

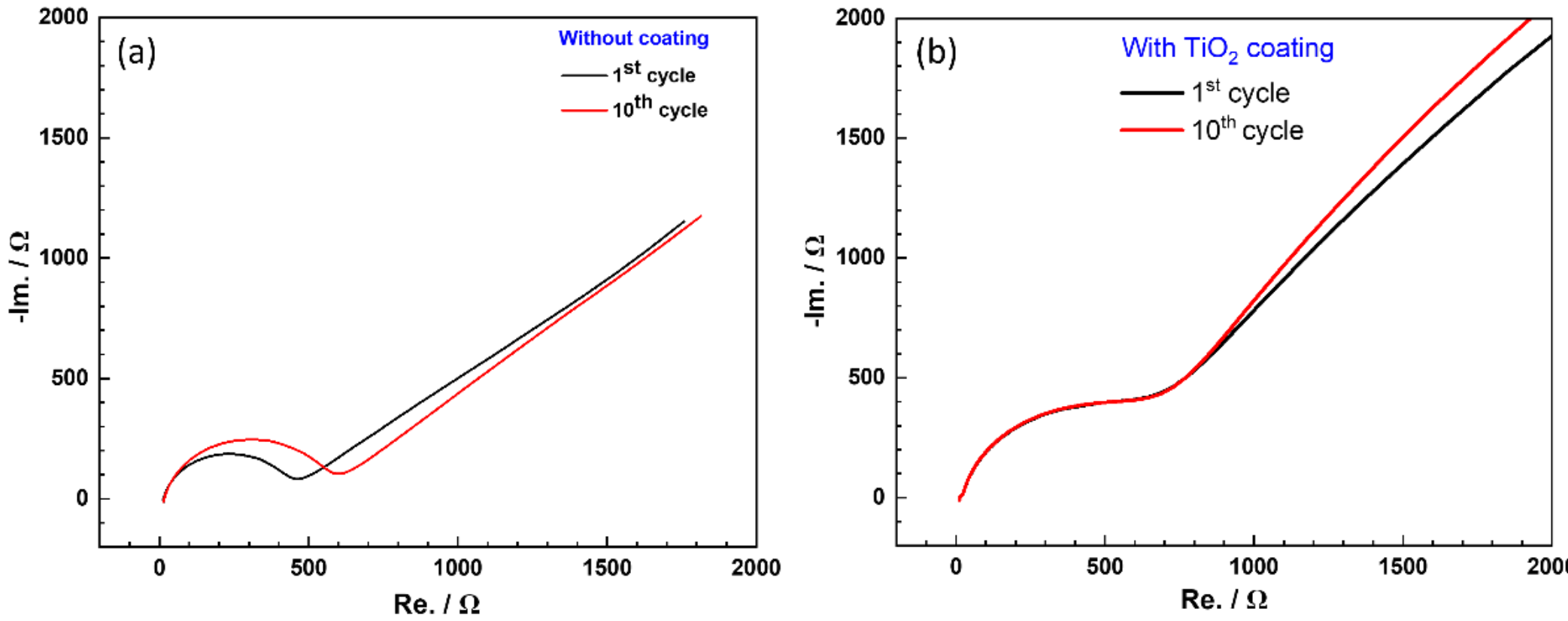

Fig. S4-8 EIS spectrum for uncoated- and 10 nm $TiO_2$ coated-sample at 25 °C (composite 80 wt. % Na-β˝-$Al_2O_3$-20 wt. % graphite before electrochemical Na storage).